\documentclass{pnastwo}

\usepackage[dvips]{graphicx}
\usepackage{pnastwoF}
\usepackage{amssymb,amsfonts,amsmath}
\usepackage{color}

\newcommand{\ch}{\textit{Chandra}}
\newcommand{\xmm}{\textit{XMM-Newton}}
\newcommand{\ofo}{0509$-$67.5}

\newcommand{\apj}{ApJ}
\newcommand{\aap}{A\&A}
\newcommand{\apjl}{ApJ}
\newcommand{\nat}{Nature}
\newcommand{\mnras}{MNRAS}
\newcommand{\pasj}{PASJ}
\newcommand{\physrep}{Phys. Rep.}
\newcommand{\araa}{ARA\&A}
\newcommand{\iaucirc}{IAU Circ.}
\newcommand{\aj}{AJ}

\url{www.pnas.org/cgi/doi/10.1073/pnas.0709640104}
\copyrightyear{2008}
\issuedate{Issue Date}
\volume{Volume}
\issuenumber{Issue Number}
\begin{document}

\title{X-Ray Studies of Supernova Remnants: A Different View of Supernova Explosions}

\author{Carles Badenes\affil{1}{Benoziyo Center for Astrophysics, Weizmann Institute of Science, Israel; and School of
    Physics and Astronomy, Tel-Aviv University, Israel}}

\contributor{Submitted to Proceedings of the National Academy of Sciences
of the United States of America}

\maketitle

\begin{article}

\begin{abstract}
  The unprecedented spatial and spectral resolutions of \ch\ have revolutionized our view of the X-ray emission from
  supernova remnants. The excellent data sets accumulated on young, ejecta dominated objects like Cas A or Tycho present
  a unique opportunity to study at the same time the chemical and physical structure of the explosion debris and the
  characteristics of the circumstellar medium sculpted by the progenitor before the explosion. Supernova remnants can
  thus put strong constraints on fundamental aspects of both supernova explosion physics and stellar evolution scenarios
  for supernova progenitors. This view of the supernova phenomenon is completely independent of, and complementary to,
  the study of distant extragalactic supernovae at optical wavelenghts. The calibration of these two techniques has
  recently become possible thanks to the detection and spectroscopic follow-up of supernova light echoes. In this paper,
  I will review the most relevant results on supernova remnants obtained during the first decade of \ch, and the impact
  that these results have had on open issues in supernova research.
\end{abstract}


\abbreviations{CC, core collapse; CSM, circumstellar medium; HD, hydrodynamic; ISM, interstellar medium; NEI,
  nonequilibrium ionization; SN, supernova; SNR, supernova remnant; WD white dwarf}

\dropcap{T}he vast majority of the hundreds of SNe discovered each year at optical wavelenghts explode very far
away. High quality light curves and spectra of some of these SNe are obtained routinely from ground-based observatories,
but the large distances involved restrict the view of the SN ejecta to a single line of sight through the receding
photosphere, and usually preclude a detailed study of the immediate surroundings of the exploding star. Only the
occasional nearby objects like SN 1987A allow us to probe the explosion and its surroundings with better resolution,
advancing our knowledge more effectively than many dozens of distant SNe.

The high quality X-ray observations of young ejecta-dominated SNRs performed by \ch\ and \xmm\ over the last decade have
presented us with a different view of the SN phenomenon, a view that to a large extent complements the shortcomings of
the optical studies of SNe. After the light from the SN fades away, the ejecta expand and cool down until their density
becomes comparable to that of the surrounding material, either the ISM or a more or less extended CSM modified by the SN
progenitor. At this point, a double shock structure is formed and the SNR phase itself begins. The forward shock or
blast wave moves into the ambient medium, while the reverse shock moves into the SN ejecta.  Because the ambient
densities are low ($n\sim 1\, \mathrm{cm^{-3}}$ or $\rho \sim 10^{-24}\,\mathrm{g\,cm^{-3}}$), the shocks are fast
enough ($v\gtrsim 1000\, \mathrm{km \, s^{-1}}$) to heat the material to X-ray emitting temperatures. These thermal
X-rays from the high-Z elements synthesized in the SN explosion are usually the dominant component in the X-ray spectra
of young SNRs.

The ability of \ch\ to perform spatially resolved spectroscopy on arcsecond scales reveals the structure of this
shock-heated SN ejecta in Galactic SNRs with an unprecedented level of detail. Because the morphology and spectral
properties of the SNRs also provide important clues about the structure of the ISM or CSM that is interacting with the
ejecta, young SNRs are the ideal setting to study simultaneously the aftermath of a SN explosion and the events leading
up to it. Understandably, the \ch\ observations of SNRs have drawn a great deal of attention from the SN community,
becoming a prime benchmark for different aspects of SN theory, in particular the state-of-the-art three-dimensional
models of CC and Type Ia SN explosions. However, the interpretation of these exceptional data sets is far from being a
trivial task.

In this review, I summarize the main results relevant to SN research obtained from SNR observations by \ch\ and other
X-ray satellites, point out the current challenges in the field, and discuss future prospects to meet these
challenges. The unique capabilities of \ch\ have obviously led to substantial progress in many other aspects of SNR
research, most notably the physics of cosmic ray acceleration in SNR shocks
\cite{decourchelle00:cr-thermalxray,warren05:Tycho,cassam-chenai07:tycho,helder09:CR_Acceleration_RCW86}. These issues
are outside the scope of the present work, which focuses on SNRs as a tool to study SN explosions. For brevity, I will
discuss almost exclusively the so-called `historical' SNRs, young objects whose ages are known well enough to model
their dynamics with a certain degree of confidence. I will also concentrate on results from non-dispersive X-ray
spectroscopy, leaving out SNR observations conducted with the high resolution gratings onboard \ch\ and \xmm.

\section{The \ch\ View of Young SNRs}

\subsection{The Power of the Data}

The excellent quality of the data generated by \ch\ for bright SNRs in our Galaxy is showcased by the false color images
that have become perhaps the mission's most recognizable and widespread visual result (Figure 1). For a SNR
with an angular diameter of $6 ^\prime$ like Cas A, \ch\ can resolve more than $10^{5}$ individual regions. Deep
exposures of bright objects usually provide enough photon statistics to obtain a good spectrum from most of these
regions at the moderate spectral resolution of the ACIS CCD detectors ($E / \Delta E \approx 10-60$). This usually
allows to detect K-shell emission from abundant elements like O, Si, S, Ar, Ca, and Fe.

\begin{figure*}[t]
  \begin{center}
    \includegraphics[width=\textwidth]{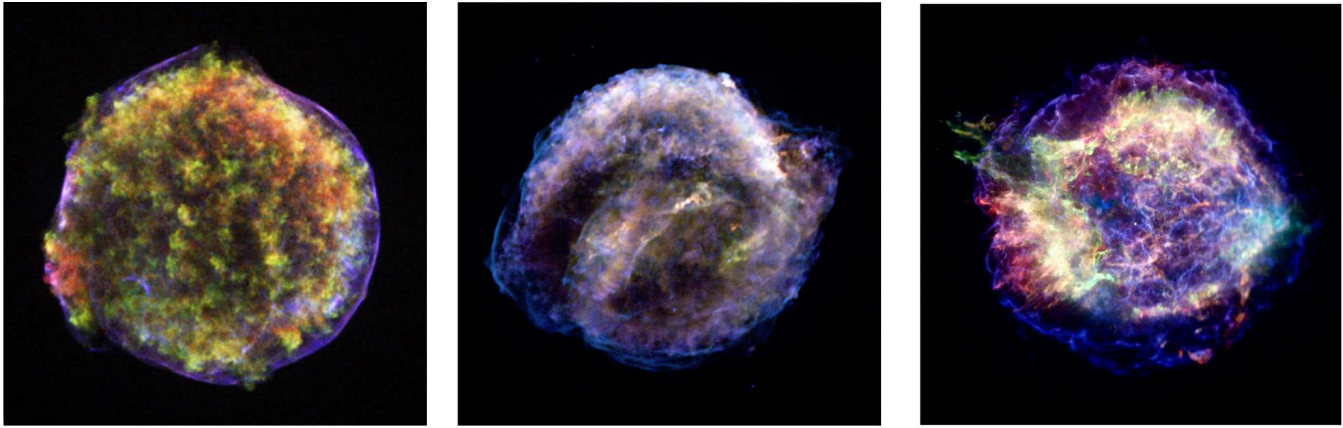}
    \caption{Three-color images generated from deep \ch\ exposures of the Tycho (left), Kepler (center), and Cas A
      (right) SNRs. The details vary for each image, but red usually corresponds to low energy X-rays around the Fe-L
      complex ($\sim$ 1 keV and below), green to mid-energy X-rays around the Si K blend ($\sim$ 2 keV), and blue to
      high energy X-rays in the 4-6 keV continuum bewteen the Ca K and Fe K blends. Images are not to scale: Tycho is
      $\sim8\, ^\prime$ in diameter, Kepler is $\sim4\, ^\prime$, and Cas A is $\sim6\, ^\prime$. Total exposure times are
      $150$, $750$, and $1000$ ks. Images courtesy of the \ch\ X-ray Center; data originally published in
      \cite{warren05:Tycho}, \cite{reynolds07:kepler}, and \cite{hwang04:CasA_VLP}.}\label{fig-1}
  \end{center}
\end{figure*}

\subsection{Basic Concepts of NEI Plasmas}

The density of the plasma inside SNRs is low enough for the ages of young objects like Cas A or Tycho to be smaller than
the ionization equilibrium timescale. The X-ray emitting plasma heated by the shocks is therefore in a state of
nonequilibrium ionization, or NEI \cite{itoh77:SNRs_NEI}. This means that the ionization state of any given fluid
element is determined not only by its electron temperature $T_{e}$, but also by its ionization timescale $n_{e}t$, where
$n_{e}$ is the electron number density and $t$ is the time since shock passage. The thermal X-ray spectrum from a young
SNR is thus intimately related to its dynamic evolution through the individual densities and shock passage times of each
fluid element. This has important implications for the ejecta emission, because of the large differences in chemical
composition across the SN debris, and the fact that the electron pool is completely dominated by the contributions from
high-Z elements, making $n_{e}$ a strong function of the ionization state \cite{hamilton84:ejecta}.

Under these circumstances, the quantitative analysis of X-ray spectra from young SNRs becomes a challenging endeavor. In
order to derive magnitudes that are relevant to SN physics, like the kinetic energy $E_{k}$ or the ejected mass of each
chemical element, it is necessary to build a hydrodynamic model of the entire SNR. One must know when each fluid element
was shocked, what its chemical composition is, how much of the SN ejecta is still unshocked at the present time,
etc. Individual fitting of each magnitude becomes impractical, because they are all related to each other, and in order
to understand the X-ray spectrum of a particular SNR, one has to understand of the object as a whole. Further
complications stem from two sources. One is a technical, but important, problem: the uneven quality of atomic data in
X-ray emission codes for NEI plasmas \cite{borkowski01:sedov,badenes06:tycho}. The other is of a more fundamental
nature: the large uncertainties in the physics of collisionless shocks, in particular the amount of ion-electron
temperature equilibration at the shock transition \cite{ghavamian07:shock_equilibration,heng09:balmer_shocks} and the
impact of cosmic ray acceleration on the dynamics of the plasma \cite{decourchelle00:cr-thermalxray}.

Most of the analysis of the X-ray emission from SNRs is done following one of two radically different approaches. The
first approach is to forgo the complex interaction between SNR dynamics and X-ray spectrum, and fit the emission of the
entire SNR or individual regions using ready-made spectral models. The most widely used are plane-parallel shock NEI
models \cite{hughes00:E0102,borkowski01:sedov}, which fit the spectra by varying $T_{e}$, some parameter related to
$n_{e}t$, and a set of chemical abundances. This has the advantages of simplicity and flexibility, but the number of
free parameters is large, the quality of the fits is often poor, and the results are hard to interpret in the framework
of SN physics and progenitor scenarios. The other approach is to model the full HD evolution of the SNR \textit{ab
  initio}, starting from a grid of SN explosion models and ISM or CSM configurations, calculate the NEI processes in the
shocked plasma, and produce a set of synthetic X-ray spectra that can then be compared to the observations. This
approach is less flexible, and usually does not allow for spectral fits in the usual sense (i.e., based on the $\chi^2$
statistic), but it is considerably more powerful in that the observations and the physical scenarios that are being
tested can be compared in a more direct way. The first HD+NEI models for SNRs were built to analyze the data from early
X-ray missions like \textit{Einstein} and \textit{EXOSAT} (e.g,
\cite{hughes85:nei,hamilton86:SN1006,borkowski94:kepler}). Modern efforts produced for \ch\ and \xmm\ data can be found
in \cite{badenes03:xray,badenes05:xray} and \cite{sorokina04:typeIasnrs}.

\section{Type Ia SNe and Their SNRs}

\subsection{Open Issues in Type Ia SNe}

Type Ia SNe are believed to be the thermonuclear explosion of a C+O WD that is destabilized when its mass becomes close
to the Chandrasekhar limit by accretion of material from a binary companion. After the central regions ignite, the
burning front propagates outwards, consuming the entire star and leading to a characteristic ejecta structure, with
$\sim 0.7\,\mathrm{M_{\odot}}$ of Fe-peak nuclei (mostly the $^{56}$Ni that powers the light curve) in the inside and
about an equal amount of intermediate mass elements (mostly Si, S, Ar, and Ca) in the outside. The amount of $^{56}$Ni
can be as high as $\sim 1\,\mathrm{M_{\odot}}$ in the brightest Type Ia SNe, and as low as $\sim
0.3\,\mathrm{M_{\odot}}$ in the dimmest Type Ia SNe, but the large scale stratification seems to be retained in all cases
\cite{mazzali07:zorro}. State-of-the-art three-dimensional explosion models still cannot reproduce these basic features
in a self-consistent way, mostly because hydrodynamic instabilities tend to destroy the stratification of nucleosynthetic
products on very short timescales \cite{kasen09:SNIa_diversity}. This relatively simple structure of Type Ia SN ejecta
is responsible for the uniformity of the light curves and spectra that makes them useful as distance indicators for
cosmology. The still poorly understood details of the multi-dimensional explosion mechanism might provide the diversity
within the uniformity, leading to bright and dim events \cite{kasen09:SNIa_diversity}. The fact that bright Ia SNe
appear to explode preferentially in star-forming galaxies \cite{gallagher05:chemistry_SFR_SNIa_hosts} suggests some
connection between the properties of the progenitors and the explosion physics, but since the identity of the
progenitors has never been clearly established, this connection remains mysterious
\cite{maoz08:fraction_intermediate_stars_Ia_progenitors}. Depending on the nature of the WD companion, SN Ia progenitors
are divided into single degenerate (the companion is a normal star, \cite{hachisu96:progenitors}) and double degenerate
(the companion is another WD \cite{iben84:typeIsn,webbink84:DDWD_Ia_progenitors}). Finding clear evidence supporting one
of these two scenarios has become a central problem in stellar astrophysics, because unknown evolutionary effects
associated with the progenitors may introduce systematic trends that could limit the precision of cosmological
measurements based on SN Ia \cite{howell09:SN_white_paper}.

\subsection{Type Ia SNRs}

Most (but not all) Type Ia SNRs reflect the relative uniformity and simplicity of their birth events. Even a cursory
glance at the \ch\ images of Tycho (Figure 1, \cite{warren05:Tycho}) or SN1006 \cite{cassam-chenai08:SN1006} reveals
strikingly symmetrical objects, without large anisotropies in the ejecta emission, in marked contrast to the turbulent
nature of CC SNRs. The morphology and X-ray spectra of SNRs with known ages and a well-established Type Ia
classification also indicate that the progenitors do not modify their surroundings in a strong way - in particular,
there is no evidence for large wind-blown cavities around the explosion sites \cite{badenes07:outflows}. This is
relevant because the current paradigm for Type Ia progenitors in the single degenerate channel predicts fast, optically
thick outflows from the WD surface \cite{hachisu96:progenitors} that would leave behind such cavities. We have to
conlcude that either most Type Ia SNe do not have single degenerate progenitors, or these fast outflows are not always
present. In fact, the dynamical and spectral properties of young Type Ia SNRs like Tycho or \ofo\ are consistent with an
interaction with the warm phase of the ISM \cite{badenes06:tycho,badenes08:0509}, and so is the detailed morphology of
the blast wave in objects where it can be studied with sufficient resolution, like SN1006 \cite{raymond07:SN1006_shocks}

A notable exception to this is the Kepler SNR. Although the recent deep (750 ks) \ch\ exposure of Kepler left little
doubt that the SNR is of Type Ia \cite{reynolds07:kepler}, its morphology shows a strong bilateral asymmetry (Figure
1), and optical observations reveal clear signs of an interaction with a dense, N-enriched CSM
\cite{blair91:kepler_optical}. This suggests that either the progenitor of Kepler's SN or its binary companion might
have been relatively massive, creating a bow-shock shaped CSM structure as the system lost mass and moved against the
surrounding ISM \cite{bandiera87:kepler}. This complex CSM interaction makes HD+NEI modeling of the Kepler SNR
challenging, and the estimation of the fundamental SN parameters difficult. Most of the detailed models for Kepler were
published when the SNR was widely believed to be of CC origin
\cite{bandiera87:kepler,borkowski94:kepler}. It is important that these models be revised in light of
the secure Ia origin of the SNR to determine the mass of $^{56}$Ni synthesized and the pre-explosion mass-loss rate from
the progenitor.

\begin{figure}
  \begin{center}
    \includegraphics[scale=0.87,angle=90]{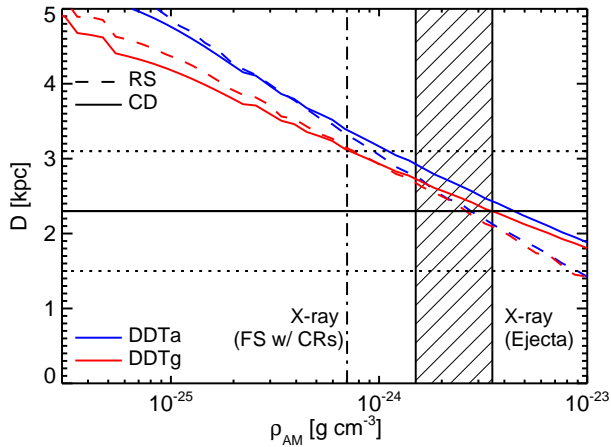}
    \caption{HD+NEI models for the Tycho SNR: Inferred distance as a function of the ambient medium
      density $\rho_{AM}$ obtained by matching the angular sizes of the reverse shock (RS, dashed plots) and contact
      discontinuity (CD, solid plots) to a bright Ia SN model (DDTa, blue plots), and a dim Ia SN model (DDTg, red
      plots). The values of $\rho_{AM}$ in the HD+NEI models of \cite{badenes06:tycho} that provide the best match to
      the X-ray emission from the SN ejecta are indicated by the striped vertical band. The value of $\rho_{AM}$
      required by the CR-modified shock models of \cite{cassam-chenai07:tycho} is indicated by a vertical dash-dotted
      line. The estimated value of the distance $D$ from \cite{smith91:six_balmer_snrs} is indicated by the horizontal
      solid line, with the dotted horizontal lines marking the upper and lower limits.}\label{fig-2}
  \end{center}
\end{figure}

The HD+NEI modeling approach has had more success in objects with simpler dynamics, like the Tycho SNR. One-dimensional
HD+NEI models with a uniform ambient medium can reproduce the fundamental features of both the dynamics and the
integrated X-ray spectrum of Tycho, provided that the ejecta is stratified, with an Fe content similar to what would be
expected from a SN Ia of normal brightness, and the ambient medium density is $\rho_{AM} \approx 2 \times
10^{-24}\,\mathrm{g\,cm^{-3}}$ (\cite{badenes06:tycho}, Figures 2 and 3). This density estimate
merits a few comments. Although it is clear that cosmic ray acceleration has a strong impact on the dynamics of the
forward shock \cite{decourchelle00:cr-thermalxray,ellison04:hd+cr}, the fact remains that the microphysics of the
process is not well understood, and cosmic ray backreaction is usually not included in full HD+NEI models for the
ejecta emission in SNRs (see \cite{patnaude09:DSA_NEI}). Instead, the assumption is made that the dynamics of the
shocked ejecta are not severely affected by cosmic ray acceleration at the blast wave, which seems reasonable in light
of the properties of the reverse shock (see discussion in Section 3 of \cite{badenes06:tycho}). In any case, it is
important to emphasize that the value of $\rho_{AM}$ required by the ejecta emission in the Tycho SNR also reproduces
the angular sizes of the reverse shock and contact discontinuity at the correct distance to the SNR (Figure 2). This
basic validation of the underlying dynamic model is essential to the HD+NEI approach, and strengthens the confidence on
the conclusions drawn about the SN explosion.

\begin{figure}
  \begin{center}
    \includegraphics[scale=0.65]{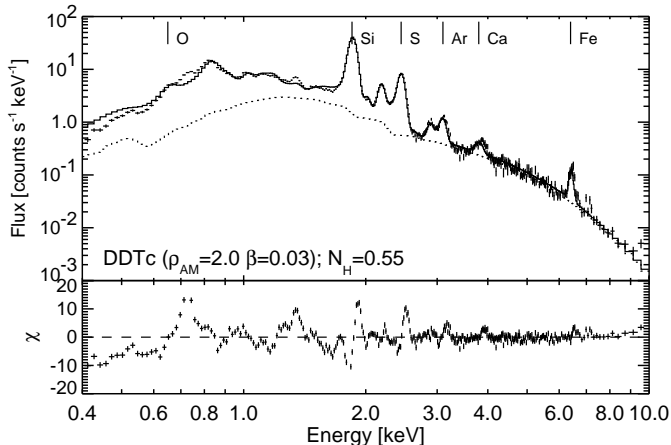}
    \caption{HD+NEI models for the Tycho SNR: Best-fit HD+NEI model to the total X-ray spectrum. The SN model is DDTc, a
      normal SN Ia model that synthesizes $0.74\,\mathrm{M_{\odot}}$ of $^{56}$Ni, interacting with an AM of $\rho_{AM}
      = 2 \times 10^{-24}\,\mathrm{g\,cm^{-3}}$. The contribution of the nonthermal emission from the blast wave is
      indicated by the dotted plot, the rest is thermal emission from the shocked ejecta (figure from
      \cite{badenes06:tycho}).}\label{fig-3}
  \end{center}
\end{figure}

The stratification of the ejecta in the Tycho SNR is required by the fact that the $n_{e}t$ of Fe in the X-ray spectrum
is lower than that of Si, which can be explained naturally if most of the Si is shocked earlier and at a higher density
than most of the Fe (see also \cite{kosenko06:Tycho}). This argument is completely independent from the constraints on
the ejecta structure obtained with optical spectroscopy of Type Ia SNe, and constitutes a clear sign that a supersonic
burning front (i.e., a detonation) must have been involved in the physics of the explosion at some stage. The most
succesful SN model found by \cite{badenes06:tycho} is a one-dimensional delayed detonation with $E_{k}=1.2
\times10^{51}$ erg that synthesizes $0.74\,\mathrm{M_{\odot}}$ of $^{56}$Ni (Figure 3) - in other words, a Type Ia SN of
normal brightness, consistent with the historical records \cite{ruiz-lapuente04:TychoSN}.

\begin{figure}
  \begin{center}
    \includegraphics[scale=1.3]{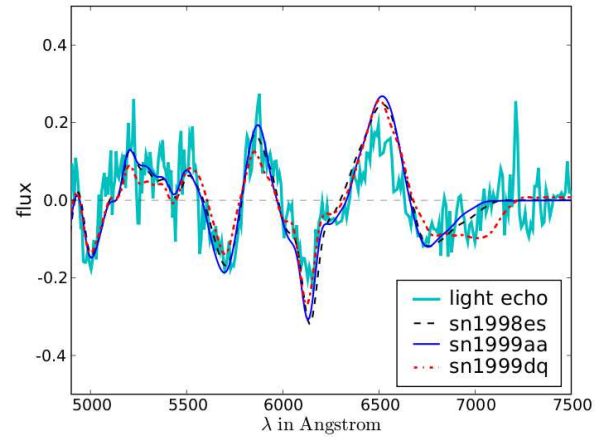}
    \caption{Direct comparison between SN and SNR spectroscopy for the LMC SNR \ofo: The LE is well matched only by spectra
      of bright Type Ia SNe like SN 1998es, SN 1999aa, and SN 1999dq (figure from \cite{rest08:0509}). }\label{fig-4}
  \end{center}
\end{figure}

One largely unexpected, but very important, development in this field has been the possibility to calibrate the results
obtained from the analysis of the ejecta emission in SNRs using the light echoes from ancient SNe. These light echoes
were originally discovered in the LMC \cite{rest05:LMC_light_echoes}, and they are important for two reasons. First,
they can provide a reliable and independent age estimate for objects that previously had none, which is a crucial
ingredient to build HD models (see e.g. \cite{badenes07:outflows}). Second, and most important, the spectroscopy of the
light echoes can be used to determine the type (CC vs. Ia) and subtype (e.g., bright Ia or dim Ia) of ancient SNe. This
technique was demonstrated by \cite{rest08:0509}, who showed that SNR \ofo\ in the LMC was originated by a bright,
$^{56}$Ni-rich Type Ia SN. At the same time, \cite{badenes08:0509} revisited the archival observations of this SNR and
found that an HD+NEI model based on a bright SN Ia explosion reproduces both the X-ray emission from the shocked ejecta
and the dynamics of the SNR (see Figures 4 and 5). This agreement between two completely independent
techniques applied to the same object has now been extended to historical SNe in our own Galaxy, with the discovery of
light echoes from Tycho and Cas A \cite{rest08:CasA_Tycho_LEs}. Spectroscopic analysis of the light echo from Tycho
\cite{krause08:tycho} confirmed that SN1572 was indeed a Type Ia SN of normal brightness, a spectacular validation of
the previous result obtained with HD+NEI models of the SNR \cite{badenes06:tycho}.

\begin{figure}
  \begin{center}
    \includegraphics[scale=0.87,angle=90]{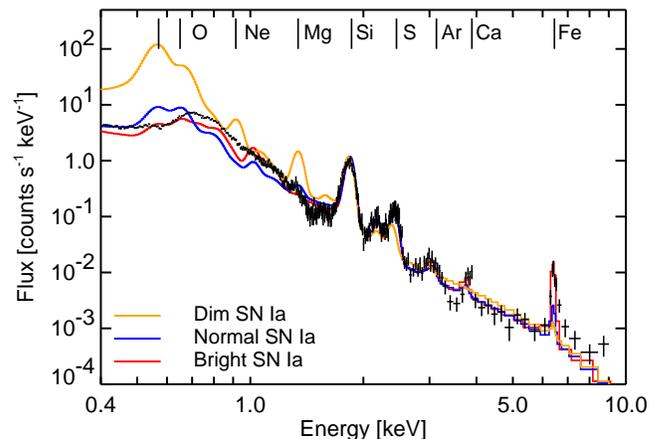}
    \caption{Direct comparison between SN and SNR spectroscopy for the LMC SNR \ofo: The X-ray spectrum of the SNR can
      only be reproduced with a bright Type Ia SN models that synthesizes $0.97\,\mathrm{M_{\odot}}$ of $^{56}$Ni
      (figure adapted from \cite{badenes08:0509}).}\label{fig-5}
  \end{center}
\end{figure}

The emergence of the X-ray observations of SNRs as well-established probes for Type Ia SNe has important implications
for SN research. Because SNRs are nearby objects, the context of the exploded star can be studied in great detail, and
relationships between specific explosion properties and progenitor scenarios can be tested \textit{in situ}. One
powerful constraint on the single degenerate scenario for SN Ia progenitors, for example, is the presence of the donor
star, which should survive the explosion \cite{canal01:companions}. So far, the search for this star in the Tycho SNR
has produced inconclusive results \cite{ruiz-lapuente04:Tycho_Binary,hernandez09:Tycho_G,kerzendorf09:Tycho_G}, but
further work in this and other SNRs should eventually clarify the identity of the progenitor of at least one SN Ia
explosion with known properties. Another recently opened line of research is the possibility to measure the metallicity
of SN Ia progenitors \textit{directly}, using Mn and Cr lines in the X-ray spectrum of their SNRs. This technique yields
a supersolar metallicity for the progenitor of Tycho \cite{badenes08:mntocr}, a measurement that will soon be extended
to other objects like Kepler. Even a small census of metallicities obtained in this way will allow us to validate the
correlations between the metallicity of the host galaxy and the SN Ia brightness found in SN surveys
\cite{cooper09:metallicity_bias_SNIa}. Finally, the SNR population in the Magellanic Clouds has great potential to
constrain Type Ia SN progenitor models, because the resolved stellar population in these nearby galaxies can be studied
in great detail. SNR \ofo, for instance, which was originated by a bright Type Ia SN, is embedded in a relatively old,
metal-poor stellar population, while SNR N103B, which has signs of a CSM interaction similar to Kepler
\cite{lewis03:N103B}, is associated with a much younger, metal-rich population \cite{badenes09:SNRs_LMC}. This suggests
that the progenitor of SNR N103B might have been relatively young and massive, and might have modified its surroundings
through some kind of mass-loss process.

\section{Core Collapse SNe and Their SNRs}

\subsection{Open Issues in Core Collapse SNe}

CC SNe mark the final events in the lives of massive ($\gtrsim 8\,\mathrm{M_{\odot}}$) stars. After nuclear fuel is
exhausted in the inner regions, the stellar core collapses to form a neutron star or a black hole. The layers of the
star that do not end up in the central compact object bounce and start propagating outwards, but at present it is
unclear exactly how the gravitational collapse of the core becomes an explosion \cite{janka07:CCSN_Review}. At the
writing of this review, the most sophisticated multi-D simulations of CC SNe still fail to explode. Although the
identity of the progenitors of CC SNe is well-established, with several identifications in pre-explosion images
\cite{smartt09:CCSN_progenitors}, the details of which specific kinds of massive stars lead to
which specific subtypes of CC SNe (e.g. Type Ib/c, Type IIL, etc.) are still under debate. Most of these uncertainties
stem from our imperfect knowledge of key processes in stellar evolution, like mass-loss mechanisms or the role played by
binarity.

\subsection{Core Collapse SNRs}

Just like Type Ia SNRs, the basic morphology of CC SNRs reflects the fundamental characteristics of their birth
events. The \ch\ images of young CC SNRs like Cas A \cite{hwang04:CasA_VLP} or SNR G292.0$+$1.8 \cite{park07:G292}
reveal a complex and turbulent structure, with marked asymmetries related both to the SN ejecta and the CSM. In contrast
to Type Ia SNRs, many CC SNRs show prominent optical emission from radiatively cooled ejecta, the hallmark of an
interaction with a relatively dense medium.

The poster child for young, ejecta-dominated CC SNRs is Cas A, an object that has played a central role in the history
of \ch. The first scientific result from the mission was the discovery of the central compact object
\cite{tananbaum99:CasA}, now thought to be a neutron star with a C atmosphere \cite{wynn09:CasA_NS_Catm}. The first
refereed publication based on \ch\ data was also devoted to Cas A \cite{hughes00:casA}. This paper discusses the
overturn of nucleosynthetic products in a large region in the SE of the SNR, where Fe-rich material can be seen clearly
ahead of Si-rich material, indicating that some of the layers deep into the onion-skin structure of the exploding star
have overtaken regions dominated by lower-Z elements (Figure 1, however, see \cite{delaney10:CasA3D} for an alternative
interpretation). Over the years, more and more has been learned about the Cas A SN and its progenitor. In 2003,
\cite{chevalier03:CasA} used the positions of the fluid discontinuities, the presence of high-velocity H, and the extent
of the clumpy photoionized pre-SN wind to infer that the Cas A explosion must have been of Type IIn or IIb SN. This
`prediction of the past', similar to the one made for the Tycho SNR, was also confirmed by the light echo, which has a
spectrum very similar to that of the Type IIb SN1993J \cite{krause08:CasA_light_echo}.

\begin{figure}[t]
  \begin{center}
    \includegraphics[width=.42\textwidth]{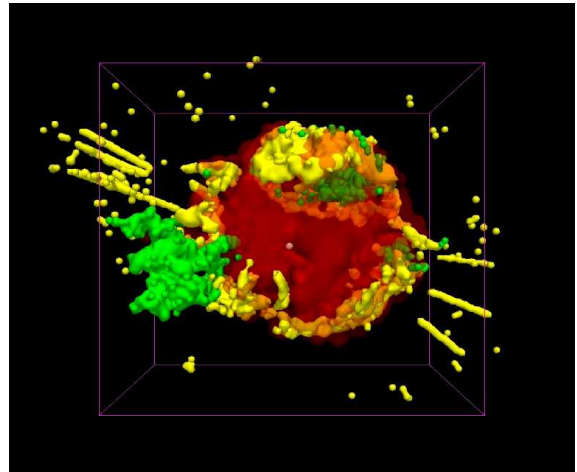}
    \caption{The Cas A SNR: Three-dimensional visualization of the distribution of ejecta,
      built by T. Delaney using Doppler shift mapping of multi-wavelength data \cite{delaney10:CasA3D}: green is
      X-ray emitting Fe; yellow is X-ray, optical and infrared emitting Ar and Si; red is infrared emitting unshocked
      ejecta; the pink dot represents the compact object. }\label{fig-6}
  \end{center}
\end{figure}

The morphology of the ejecta emission from Cas A is very rich (\cite{willingale02:CasA}, Figures 1 and 6), but
perhaps the most striking feature in the X-ray images is the jet/counterjet structure that runs from the SW to the NE
\cite{hwang04:CasA_VLP}. In order to understand the energetics of this feature and the role that it played in the
explosion, \cite{laming03:X-ray_knots_CasA,hwang03:casa-Fe} used HD+NEI simulations to interpret the fits to small X-ray
bright knots in the SN ejecta (see Figure 7). While it is clear that there was a significant local deposition of energy
around the jet region \cite{laming03:X-ray_knots_CasA}, this was probably not enought to drive the entire explosion
\cite{laming06:CasA_polar_regions} (see also \cite{wheeler08:CasA_Shape}). It turns out that the proper motion of the
compact object is not aligned with the jet axis, as one would expect in a jet-driven explosion, but perpendicular to it,
along a mysterious gap in the high-velocity ejecta revealed by \textit{HST} images of optical knots
\cite{fesen06:CasA_HST}. The expansion of these knots is consistent with an explosion date close to 1680
\cite{thorstensen01:CasA_expansion,fesen06:CasA_HST}, but the SN was not nearly as bright as other historical events
like Tycho or Kepler. Given the reddening towards the line of sight of the SNR, this suggests a low ejected mass of
$^{56}$Ni, $\sim0.1\,\mathrm{M_{\odot}}$ \cite{eriksen09:CasA}. Cas A is also one of the few CC SNRs that is young
enough to detect the decay products of $^{44}$Ti and infer an ejected mass for this isotope
($\sim10^{-4}\,\mathrm{M_{\odot}}$, \cite{vink01:CasA-44Ti}).

\begin{figure}
  \begin{center}
    \includegraphics[width=.49\textwidth]{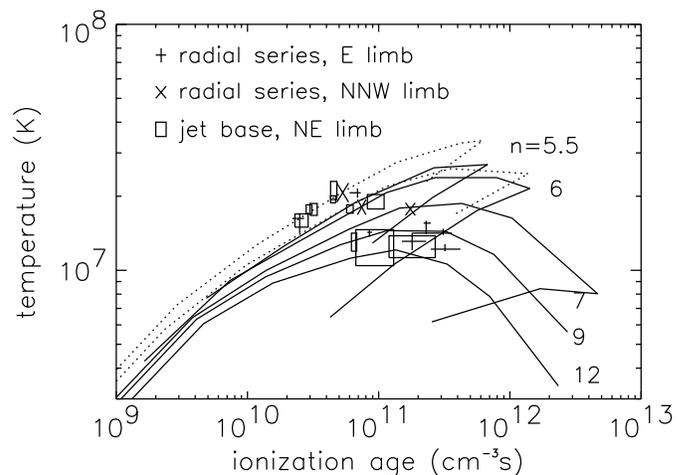}
    \caption{The Cas A SNR: Constraints on the power-law index of the ejecta structure obtained by comparing the fitted
      $T_{e}$ and $n_{e}t$ values in small X-ray knots to HD+NEI models. The lines represent models appropriate for Cas
      A, with $M_{ej}=2\,\mathrm{M_{\odot}}$, $E_{k}=2\times10^{51}$ erg and different power-law indexes. Figure from
      \cite{laming03:X-ray_knots_CasA}. }\label{fig-7}
  \end{center}
\end{figure}

Several authors have used this wealth of observational information to build detailed models for the progenitor of the
Cas A SN. According to \cite{young06:CasA_Progenitor}, all the observational constraints from the SNR can be matched by
a progenitor scenario where a 15-25 $\mathrm{M_{\odot}}$ star loses its H envelope to a binary interaction and undergoes
an energetic explosion in a `stripped' state. This view has been challenged by later works, who favor a slightly more
massive progenitor \cite{perez09:CasA_Progenitor}. A particularly controversial issue is whether the progenitor went
through a Wolf-Rayet phase before exploding, excavating a small cavity inside the dense $\rho \propto r^{-2}$ CSM that
is being overrun by the blast wave today \cite{schure08:CasA,vanveelen09:noWR_CasA,hwang09:CasA_CSM}.

This brief (and incomplete) summary of what we have learned about the Cas A SN from its SNR exemplifies the great value
of deep observations of nearby SNRs. Although the explosion itself was never observed with the conventional tools used
to study distant SNe, the birth event of the Cas A SNR might be the best-studied demise of a massive star after that of
SN 1987A.

\section{Conclusions and Future Perspectives}

This brief review on the X-ray observations of SNRs has been written with two main goals. The first is to illustrate the
power of SNR studies as probes of the SN phenomenon. The unique capabilities of modern X-ray satellites like \ch\ offer
a radically different view of the explosions and their progenitors, a vision that is independent of, and complementary
to, the conventional studies of extragalactic SNe at optical wavelengths. The X-ray data sets assembled for well-known
objects like Tycho or Cas A will be a lasting legacy of the \ch\ mission, and they represent the most detailed picture
of the structure of SN ejecta currently available at any wavelength. The recent discovery that the statistical
properties of the ejecta emission resolved by \ch\ can be used to distinguish CC from Type Ia SNRs
\cite{lopez09:typing_SNRs} is a powerful example of the potential of these data. It is therefore crucial that the
campaign of X-ray observations of SNRs with \ch\ and other satellites continue, and that deep exposures be completed for
all the objects that do not have them. The discovery of new young SNRs, while difficult, is definitely possible, as
illustrated by the $\sim100$ yr old Galactic SNR G1.9$+$0.3 \cite{reynolds09:G1.9+0.3}, and should be pursued in
parallel to the study of well-known ones.

The second goal of this review is to emphasize the importance of careful modeling for the analysis of the X-ray
spectra of SNRs. Because of the NEI character of the shocked plasma in SNRs, it is extremely difficult to interpret
their X-ray spectra in terms of magnitudes that can be used effectively to constrain explosion physics and progenitor
scenarios. A robust, quantitative analysis usually requires putting together a dynamical model for the SNR - in other
words, the X-ray emission of a SNR cannot be interpreted without understanding the object as a whole, at least to some degree. In
this context, one-dimensional HD+NEI models have been successful in recovering the fundamental properties of SN
explosions from the spatially integrated X-ray spectra of Type Ia SNRs, and validation of the technique through the
spectroscopy of SN light echoes has been possible in a few cases.

These results are certainly encouraging, but much remains to be done before the full potential of the X-ray observations
of SNRs is realized. In order to take advantage of the spatially resolved spectroscopy capabilities of \ch, it is
necessary to build multi-dimensional HD+NEI SNR models. This line of research has great promise as a benchmark for
multi-dimensional SN explosion models, which hold the key to fundamental processes like the physical mechanism
responsible for CC SNe \cite{janka07:CCSN_Review} or the origin of the diversity within Type Ia SNe
\cite{kasen09:SNIa_diversity}. 

Present and future X-ray observations of SNRs present many opportunities for SN research. These opportunities are not
devoid of challenges, but the excellent quality of the data obtained with \ch\ and other satellites fully warrant the
effort required to meet them. Because of the unique view of the SN phenomenon that they offer, SNRs should play a
central role in shaping our understanding of SN explosions.

\begin{acknowledgments}
  I want to thank Kimberly Arcand and Peter Edmonds for assistance with the \ch\ images. I also want to express my
  gratitude to all the mebers of the SN and SNR community with whom I have shared the excitement about the results of
  \ch\ and other X-ray satellites.
\end{acknowledgments}

%\bibliographystyle{pnas} 
%\bibliography{/Users/carles/Documents/mybibliography}

\end{article}

\end{document}